\documentclass[12pt,thmsa,a4paper]{article}
\usepackage{amssymb}

\usepackage{sw20lart}

\input{tcilatex2.5}
\begin{document}

\author{Ubiraci P. C. Neves and J. R. Drugowich de Fel\'{i}cio\bigskip \and {\small %
Departamento de F\'{i}sica e Matem\'{a}tica, FFCLRP, Universidade de S\~{a}o
Paulo } \and {\small Av. Bandeirantes, 3900 - CEP: 14040-901 - Rib. Preto,
SP, Brazil}}
\title{Finite-Size Scaling and Damage Spreading in Ising Systems with Multispin
Interactions }
\date{}
\maketitle

\begin{abstract}
We investigate two-dimensional Ising systems with multisping interactions of
three- ($m=3$) and four-body terms ($m=4$). The application of a new type of
finite-size algorithm of de Oliveira allow us to clearly distinguish a
first-order transition (in the $m=4$ case) from a continuous one (in the $%
m=3 $ one). We also study the damage spreading in these systems. In this
study, a dynamical phenomenon is observed to occur at a critical point
separating a chaotic phase from a frozen one. However, the width of the
interval where this transition happens does not yield a conclusive evidence
about the order of the phase transition.
\end{abstract}

\newpage

\section{Introduction}

The presence of multispin interactions can lead to severe alterations in the
behavior of statistical mechanical models such as the Baxter and the
Ashkin-Teller models. There are other models of the same type and these can
include three-, four- or more-body terms in dimension $d\geq 2$. For
instance, we may cite the Ising system with multispin interactions. For a
two-dimensional problem, the Hamiltonian $H$ of such an Ising-like system is
expressed by

\begin{equation}
-\frac{1}{k_{B}T}~H=\sum_{\langle ij\rangle }\left\{
K_{y}\,S_{i,j}\,S_{i+1,j}\,+\,K_{x}\,\prod_{l=0}^{m-1}S_{i,j+l}\right\} \;,
\label{e1}
\end{equation}
where $k_{B}$ is the Boltzmann constant, $T$ is the temperature and $%
S_{i,j}=\pm 1$ is a boolean variable localized on site $(i,j)$.

Turban and Debierre \cite{turban} have looked at this anisotropic model and
shown that it has a single transition located at the self-dual critical
point, once the above Hamiltonian is self-dual for any $m$. The critical
point can be determined through the known relation $\sinh (2K_{x})\sinh
(2K_{y})=1$, which is independent of $m$. For the isotropic case ($%
K_{x}=K_{y}$), the physical solution of this relation is $K_{x}=K_{y}=K_{c}=%
\frac{1}{2}\ln (1+\sqrt{2})=0.44068679...$ . From the spin reversal
invariance of the Hamiltonian as well as from an analysis of the energy of
domain walls, one can see that the ground state is $2^{m-1}$ times
degenerate and expect that the model will be on the same universality class
as the $q$-state Potts model whenever $q=2^{m-1}$ [1,2].

Mean-field theory \cite{turban} and finite size scaling [1,3-6] were first
used to study this problem. Further improvements of the mean-field methods 
\cite{maritan} and Monte Carlo simulations \cite{alcaraz2} helped to
indicate a first-order transition in the $m=4$ case. The method of the
fourth-order cumulant of Challa, Landau and Binder \cite{challa} and a Monte
Carlo histogram technique of Ferrenberg and Swendsen \cite{ferrenberg} were
then used to study the order of the transitions in the $m=3$ and $m=4$ cases 
\cite{drugowich}: the phase transition was characterized to be continuous in
the $m=3$ case and was asserted to be a first-order one in the $m=4$ case.

In the present study, we apply a new type of finite-size scaling algorithm 
\cite{oliveira} to study the critical properties of the above model. We
clearly distinguish a first-order transition (in the $m=4$ case) from a
continuous one (in the $m=3$ case). In addition, we obtain good collapsed
curves when the pertinent exponents are used.

We also consider the dynamics of \emph{spreading phenomena} [12,13] in the
system. A critical parameter $K_{d}$ which separates a chaotic phase from a
frozen one is obtained in both cases ($m=3$ and $m=4$). However, unlike the
finite-size scaling method, a \emph{damage spreading} analysis [14-16] seems
to be \emph{not} able to characterize the order of the phase transition.

\section{Finite-Size Scaling}

The finite-size scaling algorithm of de Oliveira [11,17] is based on two
thermodynamic quantities, namely the bulk quantity $Q$ and the surface
correlation function $\tau $.

For an Ising system, $\tau $ can be calculated as follows: one considers two
opposite surfaces of the lattice and verifies which spin state dominates
each one of them. A counter is increased by 1 if both surfaces are dominated
by the same state, otherwise it is decreased by 1. This counter is
normalized by the number of measurements and the result is denoted by $\tau
. $ In this way, $\tau $ is a step function of $T-T_{c}$. For $T<T_{c}$ the
surfaces are well-correlated and $\tau =1$, whereas for $T>T_{c}$ the
surfaces are uncorrelated and thus $\tau =0.$ This criterion holds for
second-order transitions where the correlation length diverges; however, at
a first-order transition the correlation length remains finite and thus a
multidomain state is possible even below $T_{c},$ which implies that this
normalized counter may be close to zero in some samples and close to 1 in
other samples \cite{oliveira2}.

The bulk quantity $Q$ is the average of the \emph{sign} of the sum of the
Ising spins $\pm 1$ [11,18].

Both quantities $Q$ and $\tau $ are shown [11,18] to scale as $L^{0}$ at the
critical temperature $T_{c}$ and in the thermodynamic limit ($L$ is the
linear size of the system). So they behave like Binder's fourth-order
cumulant \cite{binder}.The function $Q$ is based on a bulk measure of the
majority of spins, whereas $\tau $ measures the correlation between two
opposite surfaces of the system \cite{oliveira2}.

Another important quantity is the \emph{bulk magnetization} $M$. For an
Ising system, it is the average of the absolute value of the sum of all
spins \cite{oliveira}. At the critical temperature, this quantity scales as $%
M\sim L^{y}$ , where $y$ is the magnetic exponent. For a continuous phase
transition, the exponent is $y=D-\frac{\beta }{\nu }$, where $D$ is the
geometrical dimension of the system, $\beta $ is the exponent associated to
the spontaneous magnetization (order parameter) and $\nu $ is the thermal
critical exponent which governs the divergence of the correlation length.
The definition of $y$ can be extended even to a first-order transition,
where the exponent $\nu $ cannot be defined. In this case, $M\sim L^{D}$ and
thus $y=D$ \cite{oliveira2}.

We simulate two-dimensional Ising systems with multispin interactions and
Hamiltonian expressed by (\ref{e1}) for the cases $m=3$ and $m=4$. For each
Monte Carlo updating, spins are selected sequentially and flipped with the
thermal probability of the Metropolis algorithm. Simulations are performed
on the square lattice with periodic boundary conditions for different
lattice sizes $L$. For fixed $K_{x}$ ($=K_{y}$), we average the quantities $%
\tau $ and $M$ over a total of $(2-4)\times 10^{5}$ Monte Carlo steps after
discarding the initial $4\times 10^{4}$ transient steps. In order to obtain
these averages, we consider a subset of spins belonging to the rows of sites
which are separated from each other by multiples of $m$ lattice parameters.
In this way, we may calculate the magnetization as the average of the
modulus of $m$ times the sum of all spins belonging to those rows. The
surface correlation function may be calculated as defined above provided
that the two mentioned opposite surfaces correspond to two of those rows
which are $L/2$ lattice parameters apart from each other. The statistics is
improved by averaging over all possible subsets of rows of sites.

The bulk magnetization $M$ is measured at $K_{c}$ for different linear sizes 
$L$. The plots of $M$ versus $L$ in logarithmic scale are presented in
Figure 1 (for simulations with $m=3$) and in Figure 2 (for $m=4$). The
straight lines confirm the scaling relation $M\sim L^{y}$. For $m=3$, the
magnetic exponent obtained is $y=1.83\pm 0.03$. This result is really
consistent with the expected \emph{second-order} value $y=D-\frac{\beta }{%
\nu }=2-\frac{3}{16}=1.8125$ (this model is considered to be on the same
universality class as the $q=4$ Potts and Baxter-Wu \cite{baxter} models
with exponents $\alpha =\nu =2/3$ but $\beta =1/8$ \cite{maritan}). For $m=4$%
, we obtain the value $y=2.08\pm 0.05$ which points to the expected \emph{%
first-order} transition with magnetic exponent $y=D=2$ (remind that the $m=4$
case should be on the same universality class as the $q=8$ Potts model). So
this finite-size scaling method seems to be still very trustworthy when
applied to Ising systems with multispin interactions. It is a way to provide
a clear determination of a first-order phase transition as well as a
second-order one.

\bigskip \bigskip \bigskip

\begin{center}
\emph{Figure 1 to be inserted here.}
\end{center}

\bigskip \bigskip \bigskip

\bigskip \bigskip \bigskip

\begin{center}
\emph{Figure 2 to be inserted here.}
\end{center}

\bigskip \bigskip \bigskip

Additionally, plots of the surface correlation function $\tau $ versus $%
L^{1/\nu }~(K_{x}-K_{c})$ for different sizes $L$ can lead to estimatives of
the critical coupling $K_{c}$ and the thermal exponent $\nu $ by adjusting
these parameters so that all points collapse onto the same curve \cite
{oliveira2}. We instead prefer to show that we do obtain this collapse if
the known values of $K_{c}$ and $\nu $ are previously used. Data from
simulations with $m=3$ lead to the plot of $\tau $ versus $L^{1/\nu
}~(K_{x}-K_{c})$ (with $\nu =2/3$) as shown in Figure 3. In this graph, the
points correspond to discrete values of $K_{x}$ nearby $K_{c}$ for lattice
sizes $L=30$ (+), $60$ ($\star $) and $90$ ($\boxdot $). A good collapse is
observed for all three sets of points. In the case $m=4$ where the thermal
exponent $\nu $ is not defined we can also obtain a collapse if we consider
the artificial value $\nu =1/2$. In Figure 4, we thus plot $\tau $ versus $%
L^{2}~(K_{x}-K_{c})$ for $m=4$ and lattice sizes $L=32$ (+), $64$ ($\star $)
and $96$ ($\boxdot $). One should consider the standard deviations $\Delta
\tau $ of some points when appreciating the accuracy of the collapse: in our
simulations statistical errors may reach the value $\Delta \tau \sim 0.1$
for those points on the fall of curve corresponding to $L=96$ ($\boxdot $).

\medskip \bigskip \bigskip \bigskip

\begin{center}
\emph{Figure 3 to be inserted here.}
\end{center}

\bigskip \bigskip \bigskip

\bigskip \bigskip \bigskip

\begin{center}
\emph{Figure 4 to be inserted here.}
\end{center}

\bigskip \bigskip \bigskip

\section{Damage Spreading}

Of late years the damage spreading method has been used as a numerical
approach to study the propagation of a perturbation throughout many systems
such as spin glasses \cite{derrida}, Ising model [13-15], $q$-state Potts
model \cite{bibiano} and cellular automata \cite{derrida2}. In the present
work, we study damage spreading in Ising systems with multispin
interactions. We again investigate the Ising- like system (on the square
lattice) with a Hamiltonian described by equation (\ref{e1}) with $%
K_{y}=K_{x}$. Consider that the time evolution of two independent spin
configurations of the system is governed by the same dynamics and the same
sequence of random numbers generated in the Monte Carlo process. Let $%
A_{t}=\{S_{i,j}^{A}(t)\}$ and $B_{t}=\{S_{i,j}^{B}(t)\}$ be these two
configurations at time $t$. The \emph{total damage} $D(t)$ between them is
defined as the fraction of corresponding spins with different signs, that
is, 
\begin{equation}
D(t)=\frac{1}{2N}~\sum_{i,j}|S_{i,j}^{A}(t)-S_{i,j}^{B}(t)|\;,  \label{e2}
\end{equation}
where $N$ is the number of lattice sites.

A starting configuration is thermalized at a fixed $K_{x}$ $(\propto 1/T)$
by the Glauber dynamics. At time $t=0$, this thermalized configuration is
termed $A_{0}$ and a second configuration $B_{0}$ is created from $A_{0}$ by
flipping (``damaging'') a chosen fraction $D(0)=M/N$ of corresponding spins.
Then both configurations evolve in time according to this same dynamics and
with the same sequence of random numbers. After a transient (2000 steps per
site), we get a time average of $D(t)$ (over 7000 Monte Carlo steps). We
calculate its medium value $<D(t)>$ over several samples of configurations $%
A_{0}$ and $B_{0}$, for each $K_{x}$ and initial damage $D(0)$. From a total
of $50$ initial samples, we only average over those configurations where
damage is non null (unless all of them present a damage that has become
equal to zero).

In a Monte Carlo step at time $t$, all lattice sites $(i,j)$ are
sequentially visited and each spin $S_{i,j}(t)$ is flipped with probability 
\begin{equation}
p_{i,j}(t)=\frac{1}{1+\exp (\frac{\Delta H}{k_{B}T})}  \label{e3}
\end{equation}
where $\bigtriangleup H$ is energy change associated with such a possible
spin-flip. The numerical procedure for updating the spins consists in
generating a sequence of random numbers $r_{i,j}(t)$ uniformly distributed
in the interval $\left[ 0,1\right] $ and setting $S_{i,j}(t+1)=-S_{i,j}(t)$
if $r_{i,j}(t)\leq p_{i,j}(t)$ or setting $S_{i,j}(t+1)=S_{i,j}(t)$
otherwise. One must use the same sequence $r_{i,j}(t)$ for updating both
configurations $A_{t}$ and $B_{t}.$

We simulate Ising-like systems on square lattices $L\times L$ with periodic
boundary conditions. The averaged long-time damage $<D(t)>$ versus $K_{x}$
for the case $m=3$ (with $L=42$) is presented in Figure 5 whereas results
corresponding to $m=4$ (with $L=40$) are shown in Figure 6. For each case we
plot two curves: one of them connects the points (squares) obtained from
simulations with initial damage $D(0)=1/N=1/L^{2}$ and the other one (star
points) corresponds to $D(0)=0.90$.

\medskip \bigskip \bigskip \bigskip

\begin{center}
\emph{Figure 5 to be inserted here.}
\end{center}

\bigskip \bigskip \bigskip

\bigskip \bigskip \bigskip

\begin{center}
\emph{Figure 6 to be inserted here.}
\end{center}

\bigskip \bigskip \bigskip

In both graphs, for the lower values of $K_{x}$ (corresponding to the higher
temperatures) the long-time damage reaches the value $D^{*}=1/2$ which is
the same result from Glauber dynamics for damage spreading in the Ising
Model. In fact, there are also two states per spin in the multispin Ising
system so two corresponding spins $S_{i,j}^{A}(t)$ and $S_{i,j}^{B}(t)$
present $4$ possible configurations. For those values of $K_{x}$, all $4$
possibilities should be equally probable but only half of them would
correspond to damaged configurations. So there is a range of low values of $%
K_{x}$ where the damage does not depend on the initial conditions. However,
exactly at $K_{x}=0$, one should note $p_{i,j}(t)=1/2$ so that two
corresponding spins would (or would not) simultaneously change their states
thus preserving the initial damage ($<D(t)>=D(0)$). There is also another
situation where the initial damage is preserved: for $D(0)=1$ and $m=4$ one
can easily prove that two corresponding spins have the same probability $%
p_{i,j}(t)$ of flipping (the preservation of initial damage $D(0)=1$ has
been already observed for the $m=2$ Ising ferromagnet).

For higher values of $K_{x}$ (or lower temperatures) the damage depends on
the initial conditions. In this case, we observe that if the initial damage
is $D(0)<1/2$ then the long-time damage is $<D(t)>=0$ for both $m=3$ and $%
m=4 $. On the other hand, if $D(0)>1/2$\ then $<D(t)>\rightarrow 2/3$ for $%
m=3$ whereas $<D(t)>\rightarrow 1$ for $m=4$. This behaviour is illustrated
in Figures 5 and 6 for the chosen values $D(0)=1/L^{2}<1/2$ and $%
D(0)=0.90>1/2$.

For increasing $K_{x}$ close to $K_{c}$ the damage changes from $D^{*}=1/2$
to the value corresponding to the second plateau of each curve. Oscillations
in the critical region are due to statistical fluctuations. The standard
deviations of those points in the critical interval may vary from $\sim 0.05$
unit (star points) up to $\sim 0.15$ (squares).

Similar dynamical critical phenomenon has been already observed in the
literature for Ising \cite{moreira} and $q$-state Potts models (with
Hamiltonian given by $H=-J\sum_{<ij>}\delta (\sigma _{i},\sigma _{j})$) \cite
{bibiano}. In reference \cite{bibiano}, it has been shown that simulations
of damage spreading in the $q$-state Potts model through a Glauber dynamics
yielded a frozen phase for a temperature $T<T_{d}$ and a chaotic phase for $%
T\geq T_{d}$. Damage was observed to assume a high-temperature value $%
(q-1)/q $ above a characteristic temperature $T^{*}\geq T_{d}$. This allowed
the authors to define an interval $\bigtriangleup T=T^{*}-T_{d}$ whose $q$
dependence presented a decreasing behaviour with increasing values of $q$,
giving ``some indication of the order of the phase transition'' \cite
{bibiano}.

The above fact has motivated us to investigate the order of the phase
transiton in the $m=3$ and $m=4$ multispin Ising systems through a damage
spreading analysis. Our results above (expressed as functions of the
coupling $K_{x}$ instead of $T$) show that, in fact, there is a dynamical
critical phenomenon at a critical coupling $K_{d}$ so that for $K_{x}\leq
K_{d}$ the damage is insensitive to its initial value (\emph{chaotic phase})
whereas for $K_{x}>K_{d}$ there are two definite constant values of damage
corresponding to initial conditions $D(0)<1/2$ and $D(0)>1/2$ (\emph{frozen
phase}). We estimate $K_{d}=0.440(1)$ for $m=3$ (Figure 5) and $%
K_{d}=0.444(1)$ for $m=4$ (Figure 6). We also identify a characteristic
coupling $K^{*}\leq K_{d}$ so that for $K_{x}\leq K^{*}$ the damage assumes
the value $D^{*}=1/2$ and the curves corresponding to different initial
conditions join themselves. Then we estimate an interval $\Delta
K=K_{d}-K^{*}=0.006(2)$ for $m=3$ and $\Delta K=0.008(2)$ for $m=4$. The
first result might be compared with the interval $\Delta K\sim \frac{J}{%
2k_{B}}\left( \frac{1}{T_{d}}-\frac{1}{T^{*}}\right) \sim \frac{1}{2}\left( 
\frac{1}{0.90}-\frac{1}{0.91}\right) \sim 0.006$ (for the $q=4$ Potts model)
extracted from Fig. 3 of reference \cite{bibiano}, where the \textit{non
vanishing} interval was taken as an indicator of the continuous phase
transition. On the other hand, the estimative of the interval $\Delta K$ for 
$m=4$ \textit{does not} allow us to characterize the order of the phase
transition (which is discontinuous in this case).

\section{Conclusion}

In summary, in order to investigate the phase transition of Ising systems
with multispin interactions we have applied two methods: a finite-size
scaling algorithm of de Oliveira \cite{oliveira} and a damage spreading
analysis.

Through the finite-size scaling method we have obtained the scaling relation
for bulk magnetization, $M\sim L^{y}$, with magnetic exponents consistent
with the expected second-order transition value $y=1.8125$ (for $m=3$) and
first-order transition value $y=D=2$ (for $m=4$). In this way the orders of
phase transitions in the $m=3$ and $m=4$ multispin Ising systems have been
clearly distinguished. Additionally, plots of the surface correlation
function $\tau $ versus $L^{1/\nu }(K_{x}-K_{c})$ for different sizes $L$
have collapsed onto the same curve, showing that this technique can also be
used as a way to estimate $K_{c}$ and $\nu $.

Regarding the analysis of damage spreading in the system, we have observed a
dynamical critical phenomenon at a critical coupling $K_{d}\approx K_{c}$
separating a chaotic phase (where damage is insensitive to its initial
value) from a frozen phase (with two definite constant values of damage).
However, the width $\Delta K$ of the interval where this transition occurs
has not yielded a conclusive evidence about the order of the phase
transition.\bigskip \bigskip

\textbf{Acknowledgments}\bigskip

This work has been supported by Brazilian Agencies \textbf{FAPESP} (through
grants No. 96/5387-3 and No. 97/00918-3) and \textbf{CNPq}. We are grateful
for financial support. 

\newpage

\begin{center}
\textsl{FIGURES CAPTION}
\end{center}

\bigskip \bigskip \bigskip \bigskip

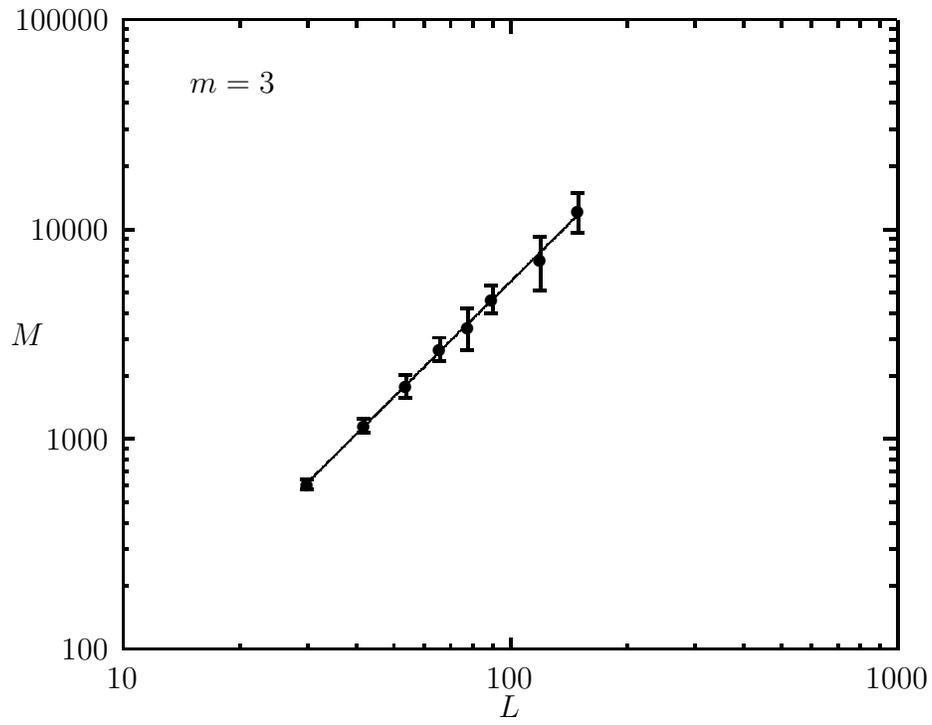
\textbf{Figure 1} - Bulk magnetization $M$ at $K_{c}$ from simulations with $%
m=3$ for linear lattice sizes $L=30,~42,~54,~66,~78,~90~,120$ and $150$. The
straight line confirms the scaling relation $M\sim L^{y}$ with magnetic
exponent $y=1.83\pm 0.03$ which is consistent with the value $y=D-\frac{%
\beta }{\nu }=1.8125$ expected for a second-order transition\emph{.}\bigskip
\bigskip

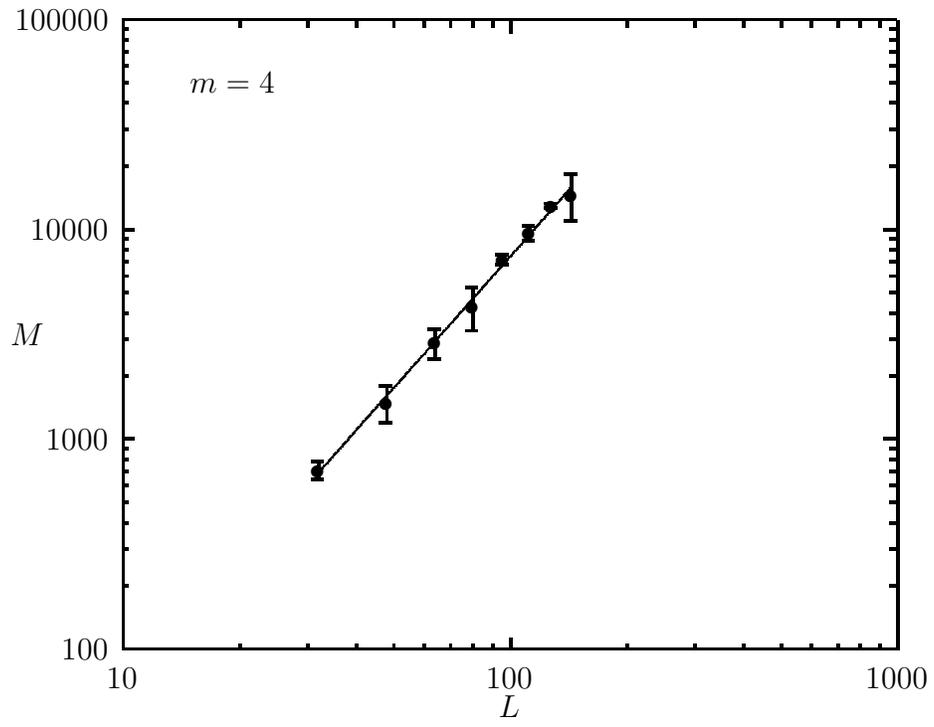
\textbf{Figure 2} - Bulk magnetization $M$ at $K_{c}$ from simulations with $%
m=4$ for linear lattice sizes $L=32,~48,~64,~80,~96,~112~,128$ and $144$.
The straight line confirms the scaling relation $M\sim L^{y}$ with magnetic
exponent $y=2.08\pm 0.05$ pointing to the expected first-order transition
value $y=D=2.$\bigskip \bigskip

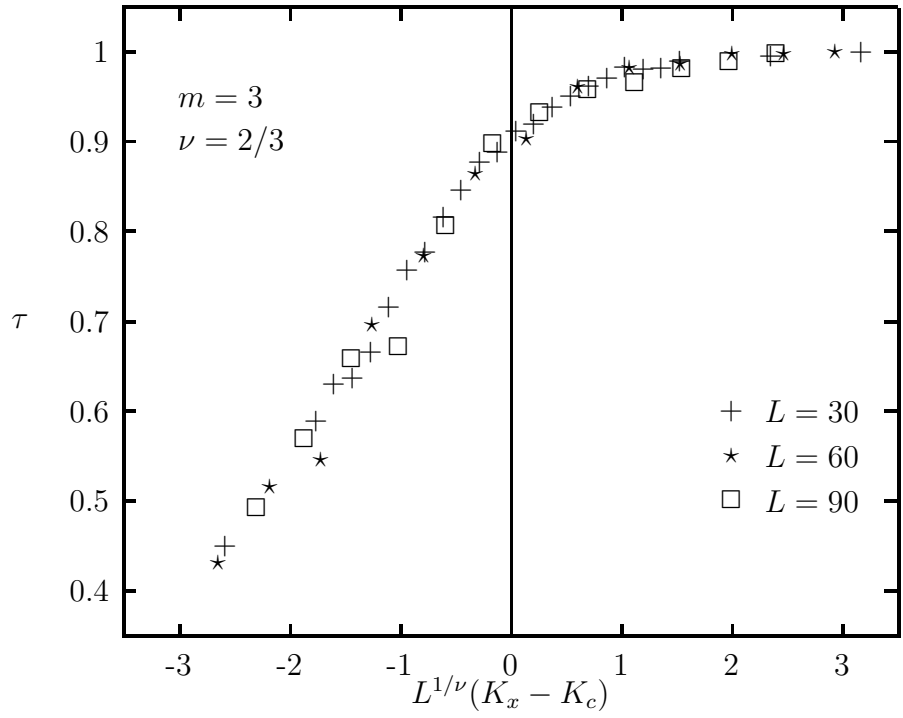
\textbf{Figure 3} - Plot of the surface correlation function $\tau $ versus $%
L^{1/\nu }~(K_{x}-K_{c})$ (with $\nu =2/3$) for $m=3$ and lattice sizes $%
L=30 $ (+), $60$ ($\star $) and $90$ ($\boxdot $).\bigskip \bigskip

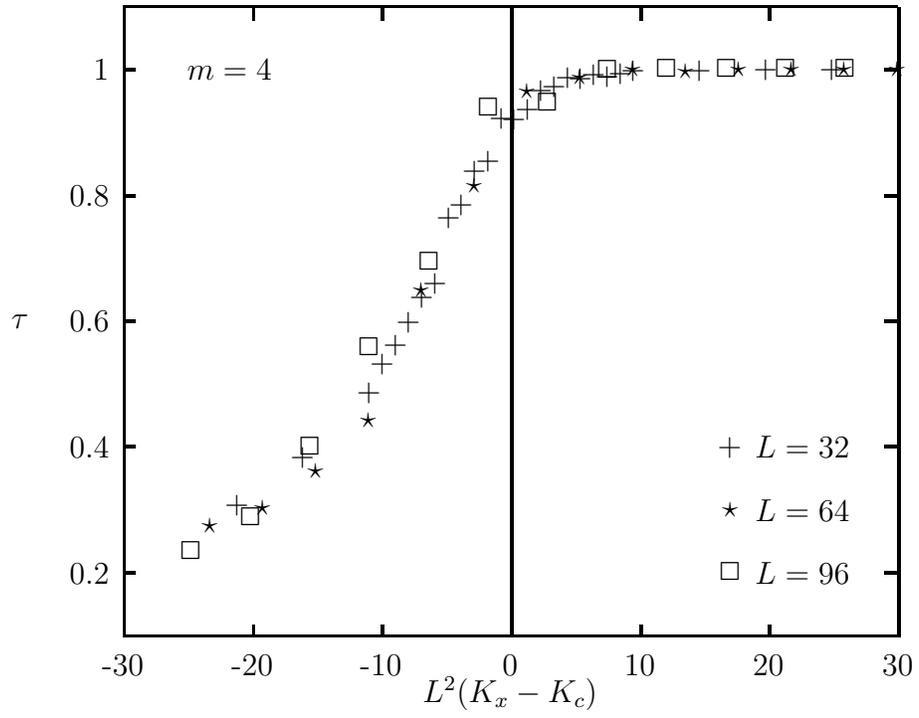
\textbf{Figure 4} - Plot of the surface correlation function $\tau $ versus $%
L^{2}~(K_{x}-K_{c})$ for $m=4$ and lattice sizes $L=32$ (+), $64$ ($\star $)
and $96$ ($\boxdot $).\bigskip \bigskip

\textbf{Figure 5 - }Average damage $<D(t)>$ versus $K_{x}$ for the $m=3$
multispin Ising system with $L=42$. The squares ($\boxdot $) represent
points obtained from simulations with initial damage $D(0)=1/N=1/L^{2}<1/2$
and the star points ($\star $) correspond to $D(0)=0.90>1/2$.\bigskip
\bigskip

\textbf{Figure 6 - }Average damage $<D(t)>$ versus $K_{x}$ for the $m=4$
multispin Ising system with $L=40$. The squares ($\boxdot $) represent
points obtained from simulations with initial damage $D(0)=1/N=1/L^{2}<1/2$
and the star points ($\star $) correspond to $D(0)=0.90>1/2$.

\newpage

\begin{center}
\textsl{FIGURE 1}\bigskip \bigskip \bigskip \bigskip

% GNUPLOT: LaTeX picture
\setlength{\unitlength}{0.240900pt} \ifx\plotpoint\undefined%
\newsavebox{\plotpoint}\fi
\sbox{\plotpoint}{\rule[-0.500pt]{1.000pt}{1.000pt}}%

\end{center}

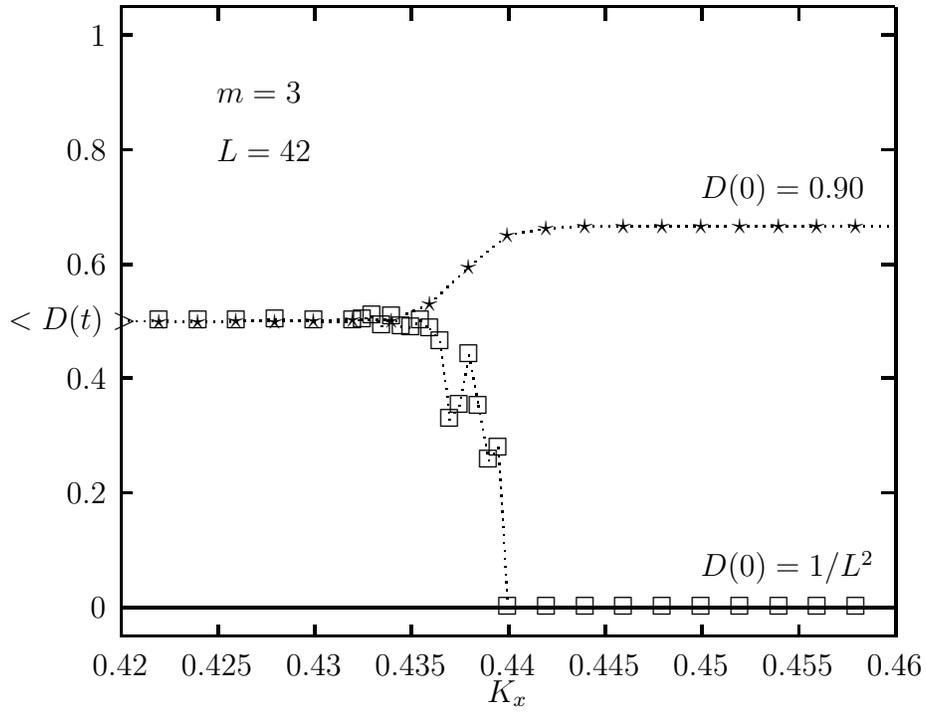
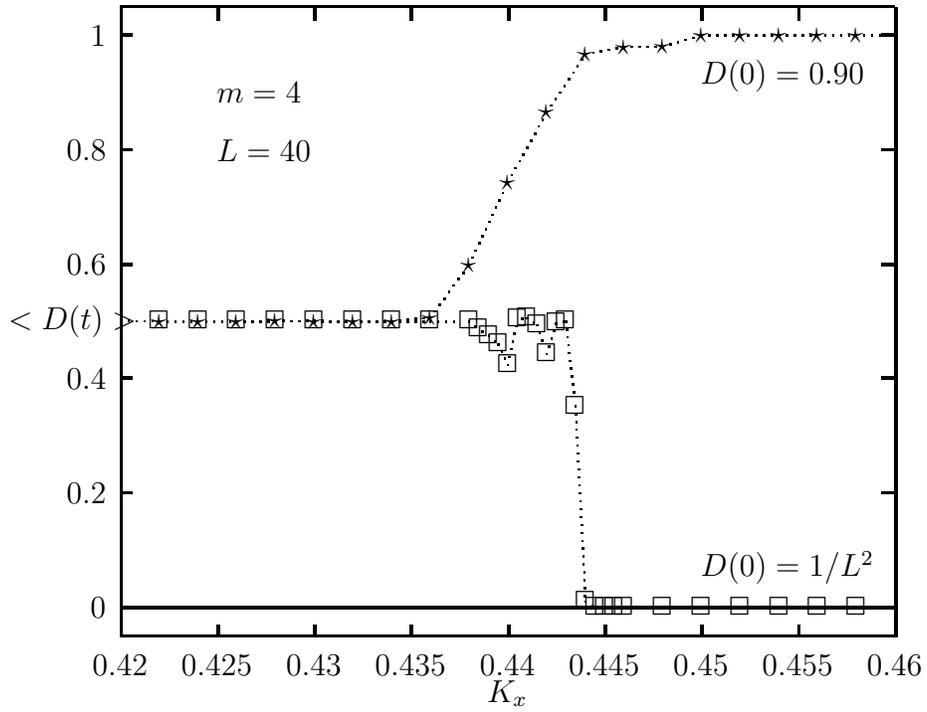
\end{document}